\documentclass[preprint,showpacs,prb,floatfix,superscriptaddress,citeautoscript
,cite]{revtex4}
\usepackage{graphicx}
\usepackage{color}
\usepackage{amsmath}

\begin{document}

\title{Strain Mapping In Single-Layer 2D Crystals Via Raman Activity}

\author{M. Yagmurcukardes}
%\email{mehmetyagmurcukardes@iyte.edu.tr}
\affiliation{Department of Physics, University of Antwerp, Groenenborgerlaan
171, B-2020 Antwerp, Belgium}

\author{C. Bacaksiz}
%\email{cihanbacaksiz@iyte.edu.tr}
\affiliation{Department of Physics, University of Antwerp, Groenenborgerlaan
171, B-2020 Antwerp, Belgium}

\author{E. Unsal}
%\email{elifunsal@iyte.edu.tr}
\affiliation{Department of  Physics, Izmir Institute of Technology, 35430, 
Izmir, Turkey}

\author{B. Akbali}
%\email{barisakbali@iyte.edu.tr}
\affiliation{Department of  Physics, Izmir Institute of Technology, 35430, 
Izmir, Turkey}

\author{R. T. Senger}
%\email{senger@iyte.edu.tr}
\affiliation{Department of Physics, Izmir Institute of Technology, 35430, 
Izmir, 
Turkey}
\affiliation{ICTP-ECAR Eurasian Center for Advanced Research, Izmir Institute 
of 
Technology, 35430, Izmir, Turkey}

\author{H. Sahin}
%\email{hasan.sahin@uantwerpen.be}
\affiliation{ICTP-ECAR Eurasian Center for Advanced Research, Izmir Institute 
of 
Technology, 35430, Izmir, Turkey}
\affiliation{Department of Photonics, Izmir Institute of Technology, 35430, 
Izmir, Turkey}


\date{\today}

\begin{abstract}
By performing density functional 
theory-based 
\textit{ab-initio} 
calculations, Raman active phonon modes of novel single-layer 
two-dimensional (2D) materials and the effect of in-plane biaxial strain on 
the 
peak frequencies and 
corresponding activities of the Raman active modes are calculated. Our findings 
confirm the Raman spectrum of 
the unstrained 2D crystals and 
provide expected variations in the Raman active 
modes of the crystals under in-plane biaxial strain. The results are 
summarized as follows; 
(i) frequencies 
of the phonon modes soften (harden) 
under 
applied tensile (compressive) strains, (ii) the response of the Raman 
activities 
to applied strain for the in-plane and out-of-plane vibrational modes have 
opposite trends, thus, 
the built-in strains in the materials can be monitored by tracking the 
relative activities of those 
modes, 
(iii) in particular, the A-peak in single-layer Si and Ge 
disappear 
under a critical tensile strain, (iv) especially in mono and diatomic 
single-layers, the shift of the peak frequencies is 
stronger indication of the strain rather than the change in Raman activities, 
(v) Raman active modes of single-layer 
ReX$_2$ (X=S, Se) are almost irresponsive to the applied strain. Strain-induced 
modifications in the Raman spectrum of 2D materials in terms of the peak 
positions and the relative Raman activities of the modes could be a convenient 
tool for characterization.

\end{abstract}

\maketitle

\section{Introduction}

The successful synthesis of graphene,\cite{Novoselov2} opened a new, famous 
field of research, 2D 
single-layer materials. 2D materials have attracted great attention due 
to their extraordinary electronic, optical, and mechanical properties that 
suit technological applications such as energy conversion, flexible electronics 
and information 
technologies.\cite{Wang1,Bhimanapati,Akinwande,Mak} Following graphene, new 2D 
single-layer structures such as 
transition-metal dichalcogenides 
(TMDs)\cite{Gordon,Coleman,Ross,Sahin2,Tongay,Horzum,Chen3,myk} 
mono-elemental 2D materials such as silicene,\cite{Cahangirov,Kara} 
germanene,\cite{Cahangirov} and group III-V binary 
compounds ($h$-BN, $h$-AlN)\cite{Sahin3,Wang2,Kim,Tsipas,Bacaksiz} were 
successfully synthesized. In addition to those in-plane isotropic 
single-layers, 2D materials with in-plane anisotropy 
such as ReS$_{2}$\cite{Tongay,Chenet,Hart,myk1}, ReSe$_{2}$\cite{Yang1,Yang2}, 
and 
black phosphorus (bp) were also widely studied\cite{bp1,bp2,bp3}. The improved 
production 
methods, such as 
chemical vapor deposition and mechanical exfoliation, enable the synthesis of 
thinner and cleaner structures. 
 
One of the most common technique 
for the characterization of a material is Raman spectroscopy\cite{raman} which 
crops information about the nature of the 
material medium entities by monitoring the characteristic vibrational energy 
levels of the structure. It also provides non-destructive analysis and 
requires 
minimum sample 
preparation. In addition, Raman measurements are also able to provide 
information about the 
substrate-free 
layer-number identification of layered 
materials,\cite{ferrari,qiao,xzhang} the 
strength of the interlayer coupling in layered materials,\cite{phtan} 
and 
interface coupling in van der Waals 
heterostructures.\cite{jbwu,jbwu2} Moreover, relative intensities of the 
Raman peaks lead to 
the determination of different-phase distributions in a 
material.\cite{colom,gouadec,havel,xzhang2}

Strain can alter electronic and vibrational 
properties of materials.\cite{guinea,khe} Raman peak 
positions and intensities strongly depend on the presence of 
strain\cite{zhni,huang} since it modifies the crystal
phonons, with stretching usually resulting in mode
softening, and the opposite for compressing. The rate of change is summarized 
in 
the Gruneissen parameters which 
determines the thermomechanical properties. Moreover, strain enhances the Raman 
spectrum of 2D materials in terms of 
the Raman intensities. Although, Raman 
spectroscopy has been widely studied in 
literature,\cite{ferrari2,berkdemir,chenet,frostig} detailed theoretical 
investigation of the strain effects on Raman peaks for 
2D materials is still limited. Here, we theoretically investigate the 
strain-dependent vibrational 
properties of several 2D single-layer
materials in terms of the peak frequencies and corresponding Raman activities.

The paper is organized as follows: Details of the computational methodology is 
given in Sec. \ref{comp}. The theory of Raman activity calculations is briefly 
explained in Sec. \ref{raman}. The evolution of peak frequencies and 
corresponding 
Raman activities under in-plane biaxial strain are discussed In Secs. 
\ref{mono}, \ref{di}, and \ref{tmd} for monoatomic, diatomic, and isotropic 
single-layer TMDs, respectively. And those for in-plane anisotropic 
single-layers of ReX$_2$ and bp are presented in Secs. \ref{rex} and \ref{bp} 
Finally, we 
conclude in Sec. \ref{Conc}.

\begin{figure*}
\includegraphics[width=18cm]{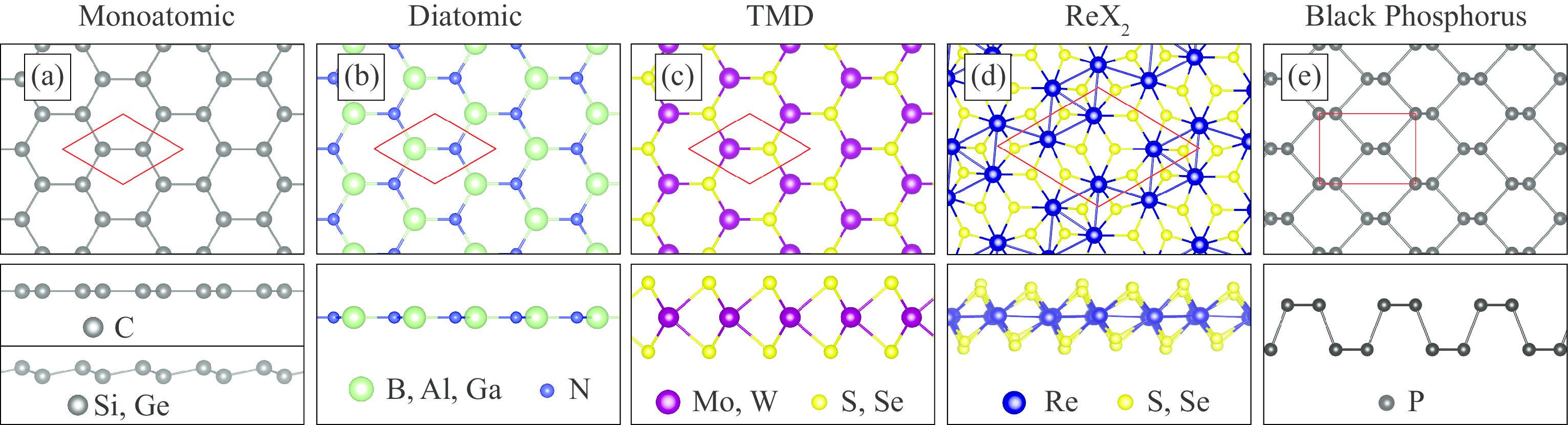}
\caption{\label{structure1}
(color line) Top and side view of single-layer crystal structures of; (a) 
monoatomic, (b) 
diatomic, (c) TMDs, (d) ReX$_2$, and (e) bp. 
Color code of individual atoms are given in each figure. Primitive cells are 
indicated 
with red solid lines.}
\end{figure*}

\section{Computational Methodology}\label{comp}

For structural optimization and vibrational properties of considered materials, 
first principle calculations 
were performed in the framework of density functional theory (DFT) as 
implemented 
in the Vienna \textit{ab-initio} 
Simulation Package (VASP).\cite{vasp1,vasp2} The Perdew-Burke-Ernzerhof 
(PBE)\cite{perdew} form of generalized gradient 
approximation (GGA) was adopted to describe electron exchange and correlation. 
The van der Waals (vdW) correction was included by using the DFT-D2 method of 
Grimme.\cite{grimme} The 
charge transfer between individual atoms was 
determined by 
the Bader technique.\cite{bader} 

The kinetic energy cut-off for plane-wave 
expansion was set to 800 eV and the energy was minimized until its variation in 
the following steps became 10$^{-8}$ eV.
The Gaussian smearing method was employed for the total energy calculations and 
the width of the smearing was chosen as 
0.05 eV. Total Hellmann-Feynman forces in the until was reduced to 10$^{-7}$ 
eV/\AA {} for the structural 
optimization. 18$\times$18$\times$1 $\Gamma$ centered \textit{k}-point 
samplings 
were used or the primitive unit cells. 
For 2$\times$2 reconstructed supercell of ReX$_{2}$ structures, 
\textit{k}-point sampling is reduced to
12$\times$12$\times$1 mesh. To avoid interaction between the adjacent 
layers, 
our calculations
were implemented with a vacuum space of 25 \AA {}.

The vibrational properties of all single-layer crystals were calculated in 
terms 
of the off-resonant Raman activities of phonon modes at the $\Gamma$ point. For 
this purpose, first the vibrational phonon modes at the $\Gamma$ point were 
calculated using finite-displacement method as implemented in VASP. Each atom in 
the primitive unit cell was initially distorted 0.01 \AA {} and the 
corresponding dynamical matrix was constructed. Then, the vibrational modes 
were 
determined by a direct diagonalization of the dynamical matrix. The $k$-point 
set was increased step by step up to 24$\times$24$\times$1 until
the convergence for the 
frequencies of acoustical modes was reached (0.0 cm$^{-1}$ for each 
acoustical 
mode). Once the accurate phonon mode frequencies were obtained, 
the change of macroscopic dielectric tensor was calculated with respect 
to each 
vibrational 
mode to get the corresponding Raman activities.

In order to investigate the response of the peak frequencies to the applied 
biaxial 
strain for Raman active modes, 
the rate of change of the peak frequency and the corresponding mode Gruneissen 
parameter were calculated. The mode Gruneissen 
parameter at any wave vector, $q$, can be calculated by the formula;

\begin{equation}\label{Gruneissen} 
\gamma(q)=-\frac{a_0}{2\omega_0(q)} \left[ 
\frac{\omega_+(q)-\omega_-(q)}{a_+-a_-} \right]
\end{equation}where $a_0$ is the relaxed (unstrained) lattice constant,
$\omega_0$(q) is the unstrained phonon frequency at wave vector $q$, 
$\omega_+$(q) and $\omega_-$(q) are the phonon
frequencies under tensile and compressive biaxial strain, respectively, and 
$a_+$ $-$ $a_-$ is the difference in the lattice 
constant
when the system is under biaxial strain. In the present study, the phonon 
frequencies are calculated 
in the $q$=0 limit, at the $\Gamma$.

%%%%%%%%%%%%%%%%%%%%%%%%%%%%%%%%%%%%%%%%%%%%%%%%%%%%%%%%%%%%%%%%%%%%%%%%%%%%%%%%
\begin{table*}
\caption{\label{main} For the single-layer crystal structures; the structure, 
planar (PL), low-buckled (LB), or puckered (P), calculated 
lattice 
parameters $a$ and $b$, the point group of the crystal, total number of 
Raman active phonon modes, calculated in-plane static (low-frequency) 
dielectric 
constant, $\epsilon_{cal}$, previously reported in-plane static dielectric 
constant, $\epsilon_{rep}$, the 
energy-band 
gap of 
the structures calculated within SOC on top of GGA 
($E_\textrm{g}^{SOC}$), for lateral orientations of the crystals; the in-plane 
stiffness, C$_x$ 
and C$_y$, and Poisson ratio, $\nu_x$ and $\nu_y$. Note : ${^*}$ The average 
in-plane static 
dielectric 
constant 
taken for anisotropic materials. }
\begin{tabular}{rcccccccccccccccc}
\hline\hline
& & &  & Point & $\#$ of & & 
  
\\
& Structure&$a$ &$b$& Group &  Raman Active & $\epsilon_{cal}$ 
&$\epsilon_{rep}$ 
& 
 $E{_g}^{SOC}$ &C$_x$ &C$_y$ &$\nu_x$ &$\nu_y$
\\
& &(\AA{})&  (\AA{}) &  & Modes &$-$ &$-$  & (eV) & (N/m)& (N/m) & $-$ & $-$ \\
\hline
Graphene& PL & 2.47 & 2.47  & D$_{3h}$  & 2 & 
23.31 & $-$ & (24$\times$10$^{-6}$)\cite{Gmitra} & 330 & 330 & 0.19 & 0.19  \\
Silicene& LB & 3.85 & 3.85 & D$_{3d}$  & 
3 
& 
8.01& $-$ & 0.001-0.01\cite{Tabert}& 54 & 54 & 0.41 & 0.41 \\
Germanene& LB & 4.01 & 4.01  &  D$_{3d}$ & 3 & 
9.04& $-$ &  0.02-0.1\cite{liu-1,liu-2}& 38 & 38 & 0.42 & 0.42 \\
\textit{h}-BN& PL & 2.51 & 2.51 & D$_{3h}$  & 
2 
&1.5 & 2-4\cite{BN} 
& 4.68(\textit{d})& 273 & 273 & 0.22 & 0.22\\
\textit{h}-AlN& PL & 3.13 & 3.13  & D$_{3h}$  & 2 & 
1.46& $-$ & 3.61(\textit{i})& 112& 112  &0.46 & 0.46 \\
\textit{h}-GaN& PL & 3.27 & 3.27 &  D$_{3h}$ & 
2 
& 1.71 & $-$ 
& 2.37(\textit{i})& 109 & 109 & 0.48& 0.48\\
MoS$_2$& 1H & 3.19 & 3.19  & D$_{3h}$  & 5 & 4.46& 4.2-7.6\cite{Huser}
& 1.56(\textit{d}) & 122 & 122  & 0.26 & 0.26\\
MoSe$_2$& 1H & 3.32 & 3.32 & D$_{3h}$  & 
5 
&5.02 &4.74\cite{Ramasubramaniam} 
& 1.33(\textit{d})& 109 & 109 & 0.25 & 0.25 \\
WS$_2$& 1H & 3.18 & 3.18  & D$_{3h}$  & 5 & 
4.12&4.13\cite{Ramasubramaniam} &  1.53(\textit{d})& 122 & 122 & 0.21 & 0.21 \\
WSe$_2$& 1H & 3.33 & 3.33 & D$_{3h}$  & 
5 
&4.67&4.63\cite{Ramasubramaniam} 
& 1.19(\textit{d})& 99 & 99 & 0.20 & 0.20 \\
ReS$_2$& 1T$^{'}$ & 6.46 & 6.38  & C$_{1h}$ & 18 & 4.18* & $-$
& 1.34(\textit{d}) & 166 & 159  & 0.19 & 0.19 \\
ReSe$_2$& 1T$^{'}$ & 6.71 & 6.60 & C$_{1h}$ & 
18 
& 4.71* & $-$
&  1.27(\textit{d}) & 138 & 138 & 0.18 & 0.18 \\
bp& P & 4.57 & 3.31 & D$_{2h}$  & 
6 
&4.32* &  $-$
&  0.88(\textit{d}) & 91 & 29 & 0.65 & 0.21\\
\hline\hline 
\end{tabular}
\end{table*}
%%%%%%%%%%%%%%%%%%%%%%%%%%%%%%%%%%%%%%%%%%%%%%%%%%%%%%%%%%%%%%%%%%%%%%%%%%%%%%%%

\begin{figure*}
\includegraphics[width=18cm]{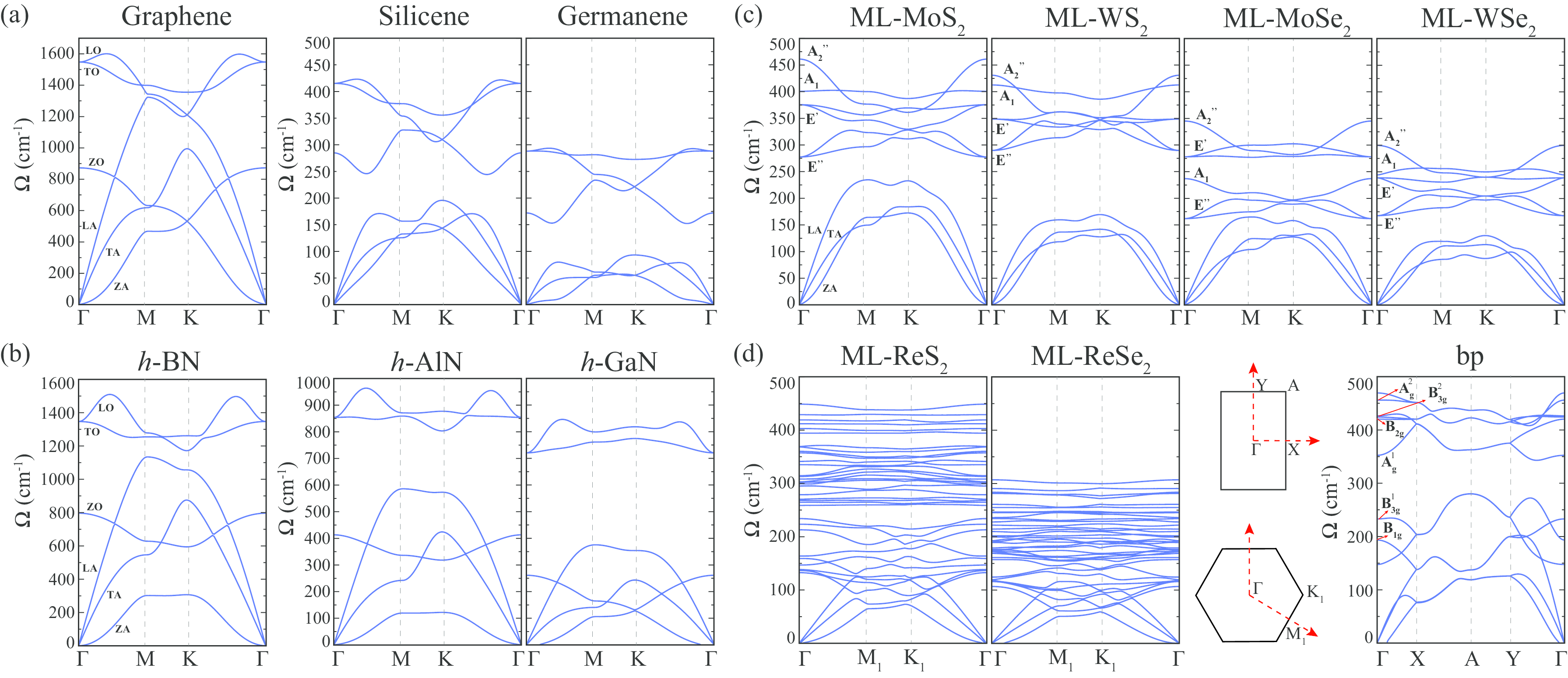}
\caption{\label{phonons}
(color line) Calculated phonon-band structures for single-layers of: (a) 
monoatomic, (b) 
diatomic, (c) isotropic TMDs, and (d) in-plane anisotropic crystals. Each 
vibrational phonon mode is named on the corresponding 
dispersion 
line. The BZ of in-plane anisotropic symmetry group is also shown. }
\end{figure*}

\section{Strain-dependent raman activity}\label{raman}

Basically in the 
Raman experiment, the sample is exposed to light and instantly scattered 
photons 
are collected. The dispersion of the collected photons with respect to shift 
in frequency 
gives the Raman spectrum. In the Raman theory, inelastically scattered 
photon originates from the oscillating dipoles of the crystal correspond 
to the Raman active vibrational modes of the crystal. 

The treatment of Raman intensities is based on Placzek's classical
theory of polarizability. According to the classical Placzek approximation, the 
activity of a Raman active phonon mode is 
proportional to $|\hat{e}_s.R.\hat{e}_i|^2$ where $\hat{e}_s$ and $\hat{e}_i$ 
stand for the polarization 
vectors of scattered radiation and incident light, respectively. $R$ is a 
3$\times$3 second rank tensor called ``Raman tensor'' 
whose elements are the derivatives of polarizability of the material with 
respect 
to 
vibrational 
normal modes,

\begin{equation}
R=\left[\begin{array}{ccc}
    \frac{
\partial\alpha_{11}}{\partial 
Q_k} & \frac{
\partial\alpha_{12}}{\partial 
Q_k} & \frac{
\partial\alpha_{13}}{\partial 
Q_k} 	\\
    \frac{
\partial\alpha_{21}}{\partial 
Q_k} & \frac{
\partial\alpha_{22}}{\partial 
Q_k} & \frac{
\partial\alpha_{23}}{\partial 
Q_k} 	\\
    \frac{
\partial\alpha_{31}}{\partial 
Q_k} & \frac{
\partial\alpha_{32}}{\partial 
Q_k} & \frac{
\partial\alpha_{33}}{\partial 
Q_k} 	\\
\end{array}\right]
\end{equation}where the $Q_k$ is the normal mode describing the whole motion of 
individual 
atoms participating to the $k^{th}$ vibrational mode and $\alpha_{ij}$ is the 
polarizability tensor of the material. The term $|\hat{e}_s.R.\hat{e}_i|^2$ is 
called the 
Raman activity which is calculated from the 
change of polarizability. For a backscattering experimental geometry, if 
orientational averaging is considered, the Raman activity 
is represented in terms of Raman invariants given by,

\begin{align}\label{invariants} 
\tilde{\alpha}_s \equiv &\frac{1}{3} 
(\tilde{\alpha}_{xx}+\tilde{\alpha}_{yy}+\tilde{\alpha}_ {zz}) \\
  \beta \equiv &\frac{1}{2} 
\{(\tilde{\alpha}_{xx}-\tilde{\alpha}_{yy})^2+(\tilde{\alpha}_{yy}-\tilde{\alpha
}
_{zz})^2+(\tilde{\alpha}_{zz}-\tilde{\alpha}_{xx})^2 \nonumber \\ 
&+6[(\tilde{\alpha}_{xy})^2+(\tilde{\alpha}_{yz})^2+(\tilde{\alpha}_{xz})^2]\} 
\end{align}where $\tilde{\alpha}_s$ and $\beta$ represent the isotropic and 
anisotropic 
parts of 
the derivative of polarizability tensor, respectively. The $\tilde{\alpha}$ 
represents the derivative of polarizability 
with respect to a normal mode. In the representation of the activity in terms 
of the variables, the activity is invariant under the orientation of sample. 
Finally, using the forms of isotropic and anisotropic polarizability 
derivative tensors, the Raman activity, R$_A$, can be written as;

\begin{equation}\label{activity-final} 
\textnormal{R}_A=45\tilde{\alpha}^2+7\beta^2
\end{equation}

In fact, every Raman active phonon mode has a finite scattering intensity which 
is measured in experiments. The term 
Raman activity, R$_A$, is directly related to the intensity through the 
following formula;

\begin{equation}\label{intensity} 
I_{Raman}=\frac{\textnormal{I}_0\pi^2}{45\varepsilon_0^2\lambda^4}\textnormal{R}
_A
\end{equation}
where I$_0$ is the intensity of incoming light, $\lambda$ is its wavelength, 
and 
R$_A$ is the Raman activity. In the rest of the present study, Raman activities 
are discussed instead of intensities since 
the Raman activity is independent of the wavelength and 
the intensity of incoming light and it is given in terms of Raman invariants (see 
Eq. \ref{activity-final}).

\begin{figure}
\includegraphics[width=7.5cm]{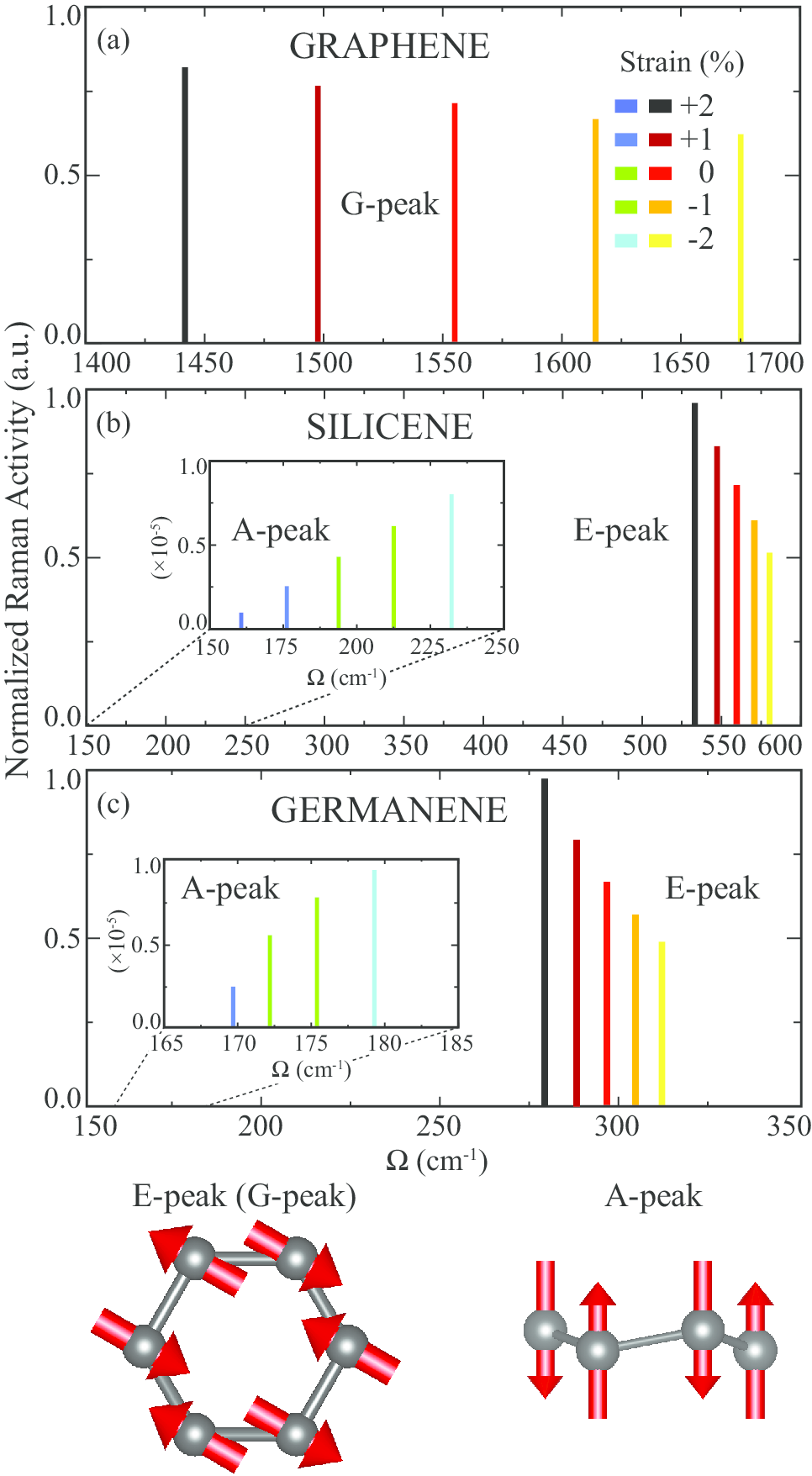}
\caption{\label{structure2}
(color line) The response of activity of Raman active modes to the applied 
biaxial strain for (a) graphene, (b) silicene, and (c) germanene. The insets in 
(b) and (c) 
are given for A-peak phonon mode.}
\end{figure}

\subsection{Mono-Atomic Single-Layer Crystals}\label{mono}

Monoatomic single-layers of graphene, silicene, and germanene have 
hexagonal crystal 
structures. Due to 
\textit{sp}$^2$
hybridization of C atoms in graphene, its structure is planar and belongs to 
$P\bar{6}/mmm$ space group symmetry. On the other hand, \textit{sp}$^3$
hybridization in silicene and germanene results in a buckled geometry (see Fig. 
\ref{structure1}(a)). The 
structure of the two buckled single-layers belong to $P\bar{3}m\bar{1}$ space 
group. 
Graphene, silicene, and germanene are known to exhibit tiny electronic band gap 
of 
24$\times$10$^{-3}$\cite{Gmitra}, 1.55-7.90 
\cite{Tabert} and 24-93 
meV\cite{liu-1,liu-2}, respectively. 

Graphene, silicene, and germanene exhibit 6 phonon 
branches that consist of 3 
acoustical and 3 optical branches (see Fig. \ref{phonons}(a)). One of the 
optical phonon modes represents out-of-plane vibrational motion of the atoms 
and 
named as A-peak phonon 
mode. The other two optical modes have in-plane vibrational characteristic and 
known as the G-peak in graphene and E-peak in 
silicene and germanene.
G- and E-peak are doubly degenerate at the $\Gamma$ due to in-plane isotropy 
of the crystals.

In the case of graphene, the frequency of A-peak is 872.8 cm$^{-1}$ and it is 
Raman 
inactive due to the planar crystal structure. On the other hand, G-peak is a 
characteristic Raman active mode in graphene and its 
frequency is 1555.0 cm$^{-1}$. When an in-plane biaxial strain is applied to 
the crystal, a 
significant phonon softening (hardening) occurs under tensile (compressive) 
strain cases which is expected due to positive 
mode Gruneissen parameter (see Table \ref{main2}). As shown in Fig. 
\ref{structure2}(a), the frequency of G-peak softens to 
1441.8 cm$^{-1}$ at 2\% of tensile strain while it hardens to 1675.1 cm$^{-1}$ 
at 2\% of compressive strain. The variation of Raman activity of G-peak is also 
shown in Fig. 
\ref{structure2}(a). It is seen that as the structure is biaxially stretched, 
the dipole between oppositely vibrating atoms gets 
larger, the polarizability increases, and hence the Raman activity increases. 
Contrarily, when the structure is 
compressed, the length of the dipole gets smaller and the Raman activity 
decreases. 

The frequency of A-peak is 193.8 cm$^{-1}$ for silicene and due to the buckled 
structure A-peak is found to be Raman active. The E-peak frequency is found at 
559.6 cm$^{-1}$ and is known to be another Raman 
active mode in silicene. Experimentally, E-peak is much 
more prominent than the A-peak due to its much higher 
Raman activity. The Raman activity of E-peak is 10$^5$ times of that of the 
A-peak. As shown in Fig. \ref{structure2}(b), when biaxial strain is applied, 
both peaks soften (harden) under tensile (compressive) strain cases. However, 
the response of Raman activities of A- and E-peak is opposite. As the structure 
is stretched, the length of dipole in A-peak gets 
smaller while it gets larger in E-peak. Thus, the polarizability hence the 
Raman activity of A-peak decreases while the it increases 
for E-peak. The situation is reversed in the compressive strain case as shown 
in Fig. \ref{structure2}(b). 
The important point is that the Raman activity of A-peak may disappear under 
high tensile strain values ($>$2\%) as the buckling 
of the structure decreases.

For the germanene, the peak position of A-peak is at 172.2 cm$^{-1}$. Since 
germanene is a softer material than silicene, the peak 
frequency of A-peak is smaller. As in the case of silicene, A-peak is Raman 
active for 
germanene due to the \textit{sp}$^3$
hybridization of Ge atoms. The frequency of E-peak is 296.8 cm$^{-1}$ which is 
also much smaller than that of in silicene. Both peaks display the same 
behavior for peak frequencies and corresponding Raman 
activities as in the case of silicene. As shown in Fig. 
\ref{structure2}(c), the E-peak displays a phonon softening to 
279.4 cm$^{-1}$ at 2\% of tensile strain while it displays a phonon hardening 
to 312.2 cm$^{-1}$ 
at 2\% of compressive strain. In addition, the Raman activity of A-peak 
decreases and disappears at 2\% of tensile strain.

Each monoatomic single-layer displays 
different responses 
to applied strain. The difference in the slope of the curves in Fig. 
\ref{structure4}(a) occurs due to different 
mode Gruneissen parameter($\gamma$) of G- and E-peak in each material. 
The $\gamma$ values are given in Table \ref{main2} and graphene has the largest 
$\gamma$ value which is a result of strong C-C bonds in the crystal. As shown 
in Fig. \ref{structure4}(b), the amount of change of 
Raman activities are also different for each single-layer crystal. Stiff 
materials, with strong interatomic bonds, charges are uniformly distributed 
between the atoms and they are not localized 
in any region even at relatively high strains. Therefore, the 
change of dielectric constant hence the 
change of Raman activity is linear. However, in the case of silicene and 
germanene the charge is localized between the atoms and thus, as the strain is 
increased the Raman activity displays a nonlinear 
change (see Fig. \ref{structure4}(b)).

\subsection{Diatomic Single-Layer Crystals}\label{di}

Similar to the crystal structure of graphene, diatomic 
single-layers of group-III Nitrides (\textit{h}-BN, 
\textit{h}-AlN, and \textit{h}-GaN) have planar, one-atom-thick structure as 
shown in 
Fig. \ref{structure1}(b). 
The crystal structures belong to space group of 
\textit{P}6$_{3}$/\textit{mmc}. Single-layers of the group-III Nitrides exhibit 
6 phonon branches (see Fig. 
\ref{phonons}(b)). In addition to the A-peak, the doubly degenerate in-plane 
phonon mode, E$_{2g}$, are the optical phonon 
modes of the single-layers. The A-peak is Raman inactive mode for the group-III 
Nitrides while E$_{2g}$ is the prominent 
peak.

\begin{figure}
\includegraphics[width=8.5cm]{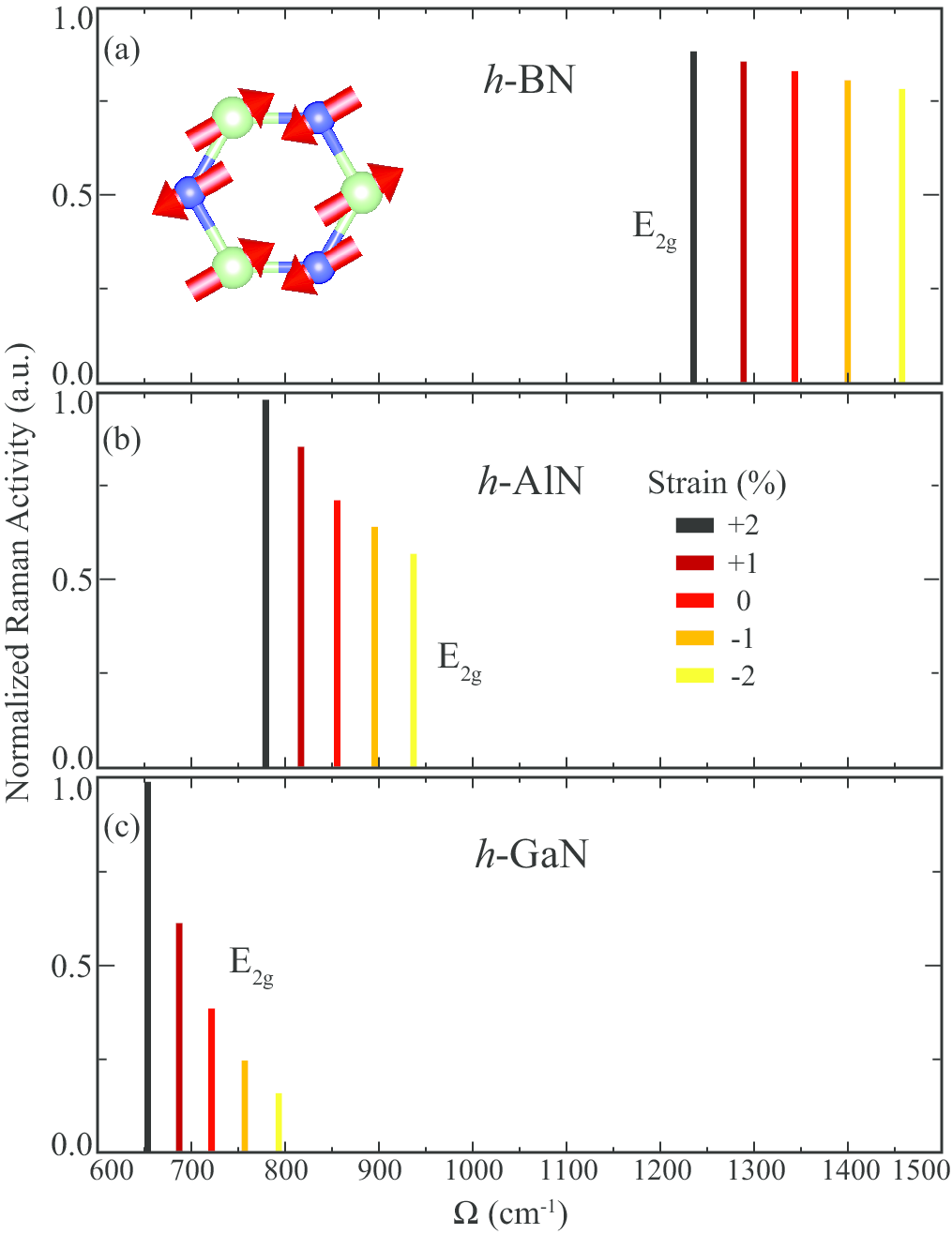}
\caption{\label{structure3}
(color line) The effect of biaxial strain on both the peak frequency and the 
corresponding Raman activity of the E$_{2g}$ phonon modes for single-layer; 
(a) 
\textit{h}-BN, (b) 
\textit{h}-AlN, and (c) 
\textit{h}-GaN. The vibrational motion of individual atoms in E$_{2g}$ phonon 
mode is shown in the inset.}
\end{figure}

The frequency of A-peak is 800.9 cm$^{-1}$ for \textit{h}-BN while the 
frequency of E$_{2g}$ is found at 1343.4 cm$^{-1}$. Due to the occupied 
in-plane orbitals in the crystal, the frequency of E$_{2g}$ is 
much greater than that of A-peak. Under biaxial strain, the E$_{2g}$ peak 
reveals a 
softening to 1235.5 cm$^{-1}$ at 2\% stretching 
and displays a hardening to 1457.7 cm$^{-1}$ at 2\% compression. The Raman 
activity of the E$_{2g}$ peak 
increases with increasing tensile strain while it decreases with increasing of 
the compressive strain as shown in Fig. 
\ref{structure3}(a). As in the case of graphene, the change of Raman activity 
is 
also linear for \textit{h}-BN due to the strong B-N bonds which preserves the 
charge distribution even at high strains. The response of Raman inactive A-peak 
to the applied strain is not discussed. 

\begin{figure}
\includegraphics[width=15cm]{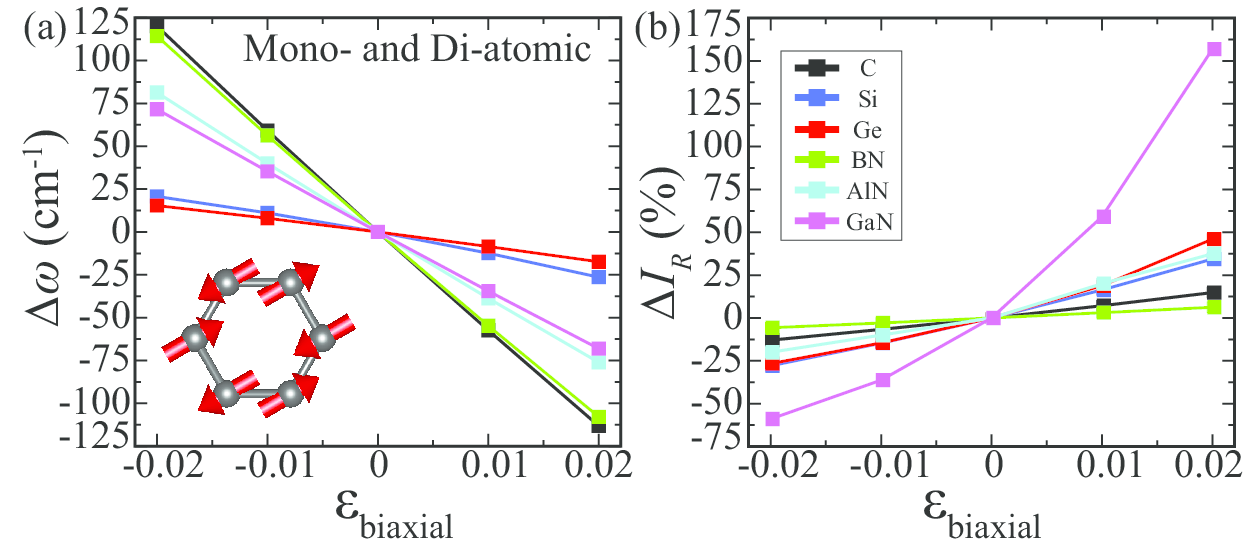}
\caption{\label{structure4}
(color line) (a) The response of peak positions and (b) corresponding Raman 
activities to the applied 
biaxial strain for single-layers of mono- and diatomic crystal structures for 
G-, E-peak, and E$_{2g}$ mode.}
\end{figure}

In the case of \textit{h}-AlN, the frequency of the A-peak softens to 412.5 
cm$^{-1}$ since the stiffness is smaller than that 
of \textit{h}-BN. The frequency of E$_{2g}$ is 855.5 cm$^{-1}$ which is also 
smaller than that of the \textit{h}-BN 
due to the same reason. When the single-layer \textit{h}-AlN is biaxially 
stretched, the peak position of E$_{2g}$ softens to 
779.4 cm$^{-1}$ while it hardens to 936.9 cm$^{-1}$. In addition, the Raman 
activity displays the same trend as in the case of 
\textit{h}-BN but the amount of change is different as shown in Fig. 
\ref{structure4}(b). The atomic bond length is much larger in \textit{h}-AlN 
than 
that of \textit{h}-BN thus, increasing strain causes nonlinear change in the 
Raman activity of \textit{h}-AlN
because of the localized charge densities in the 
crystal.

The lowest frequencies of both A-peak and the E$_{2g}$ mode are found for 
single-layer \textit{h}-GaN since it is the most flexible material through all 
considered group-III Nitrides. The peak positions of the two modes are 263.6 
and 721.6 cm$^{-1}$ for A-peak and E$_{2g}$, 
respectively. The highest rate of change of 
peak frequency with respect to the unstrained frequency is found for 
\textit{h}-GaN as given in Table \ref{main2} by the 
mode Gruneissen parameter. 

As shown in Fig. \ref{structure4}(b), the response of Raman activity of each 
material differs as the applied strain 
increases. As 
mentioned in the theory part, the Raman activity is a function of 
$\tilde{\alpha}^2$ which depends on static dielectric constant, $\epsilon$, and 
the $\epsilon$ is a 
function of square of charges on the atoms. Graphene and \textit{h}-BN display 
linear response even at $\pm$2\% strain (see 
Fig. \ref{structure4}(b)) due 
to the strong C-C and B-N bonds which preserve 
charge distribution in the crystal. As the in-plane stiffness decreases, the 
conservation of charge distribution is not possible with increasing strain 
value. Thus, silicene and germanene
exhibit nonlinear response out of the strain range $\pm$1\%. Moreover, special 
to the case of \textit{h}-GaN, the charges are 
initially localized on N atoms which indicates relatively weak Ga-N bonding in 
the crystal. Therefore, even at small strains, 
the Raman activity of \textit{h}-GaN displays nonlinear responses.  
   
%%%%%%%%%%%%%%%%%%%%%%%%%%%%%%%%%%%%%%%%%%%%%%%%%%%%%%%%%%%%%%%%%%%%%%%%%%%%%%%%
\begin{table}
\caption{\label{main2} Unstrained peak frequency, $\omega_0$, rate of change of 
peak frequency, 
$\frac{1}{\omega}\frac{d\omega}{d\varepsilon}$, corresponding 
mode Gruneissen parameter, $\gamma$, and reported value of mode 
Gruneissen parameter in previous studies, $\gamma_{rep}$, 
for 
mono- and diatomic single-layer crystals. Only the prominent and 
high-frequency 
peaks are considered.}
\begin{tabular}{lcccccccccccccccc}
\hline\hline
 & $\omega_0$ & $\frac{1}{\omega}\frac{d\omega}{d\varepsilon}$ & 
$\gamma$ & $\gamma_{rep}$
\\
 &(cm$^{-1}$) & (\%) & $-$ &$-$
\\\hline
Graphene &  1555.0 & 3.75& 1.87 & 1.85\cite{gru2} \\

Silicene &  559.6 & 2.10 & 1.05 & $-$ \\

Germanene &  296.8 & 2.78 & 1.39 &  $-$ \\

\textit{h}-BN &  1343.4 & 4.13& 2.07 & 1.70\cite{gru3} \\

\textit{h}-AlN &  855.5 & 4.60 & 2.30 & $-$ \\

\textit{h}-GaN &  721.6 & 4.84 & 2.42 & $-$ \\

\hline\hline 
\end{tabular}
\end{table}
%%%%%%%%%%%%%%%%%%%%%%%%%%%%%%%%%%%%%%%%%%%%%%%%%%%%%%%%%%%%%%%%%%%%%%%%%%%%%%%%

%%%%%%%%%%%%%%%%%%%%%%%%%%%%%%%%%%%%%%%%%%%%%%%%%%%%%%%%%%%%%%%%%%%%%%%%%%%%%%%%
\begin{table*}
\caption{\label{main3} Unstrained peak frequency, $\omega_0$, rate of change of 
peak frequency, 
$\frac{1}{\omega}\frac{d\omega}{d\varepsilon}$, corresponding 
mode Gruneissen parameter, $\gamma$, for 
in-plane isotropic single-layer TMDs. }
\begin{tabular}{lcccccccccccccccc}
\hline\hline
E$^{''}$ & $\omega_0$ & $\frac{1}{\omega}\frac{d\omega}{d\varepsilon}$ & 
$\gamma$&  
E$^{'}$ & $\omega_0$ & $\frac{1}{\omega}\frac{d\omega}{d\varepsilon}$ & 
$\gamma$  &
 A$_1$ & $\omega_0$ & $\frac{1}{\omega}\frac{d\omega}{d\varepsilon}$ & 
$\gamma$  \\
&(cm$^{-1}$) & (\%) & $-$ & \vline  &  (cm$^{-1}$) & (\%) & $-$  & \vline  & 
(cm$^{-1}$) & (\%) & $-$
\\\hline
MoS$_2$ & 277.8 & 1.04 & 0.52  & \vline  &  375.8 & 1.36 & 0.68  & \vline  & 
401.0 &0.46 & 0.23 & \\  

MoSe$_2$ &  162.2 & 0.77 & 0.39  & \vline  & 278.3 & 1.06 & 0.53  & \vline  & 
237.3 & 0.42 & 
0.21
\\

WS$_2$ &  289.5 & 0.98 & 0.49 & \vline  &  348.2 & 1.27 & 0.64  & \vline  & 
412.4 &0.51 & 
0.25   
\\

WSe$_2$ &  167.9 & 0.77 & 0.39  & \vline  &  239.0 & 1.08 & 0.54  & \vline  & 
244.4 &0.44 & 
0.22
\\
\hline\hline 
\end{tabular}
\end{table*}
%%%%%%%%%%%%%%%%%%%%%%%%%%%%%%%%%%%%%%%%%%%%%%%%%%%%%%%%%%%%%%%%%%%%%%%%%%%%%%%%

\begin{figure}
\includegraphics[width=8.5cm]{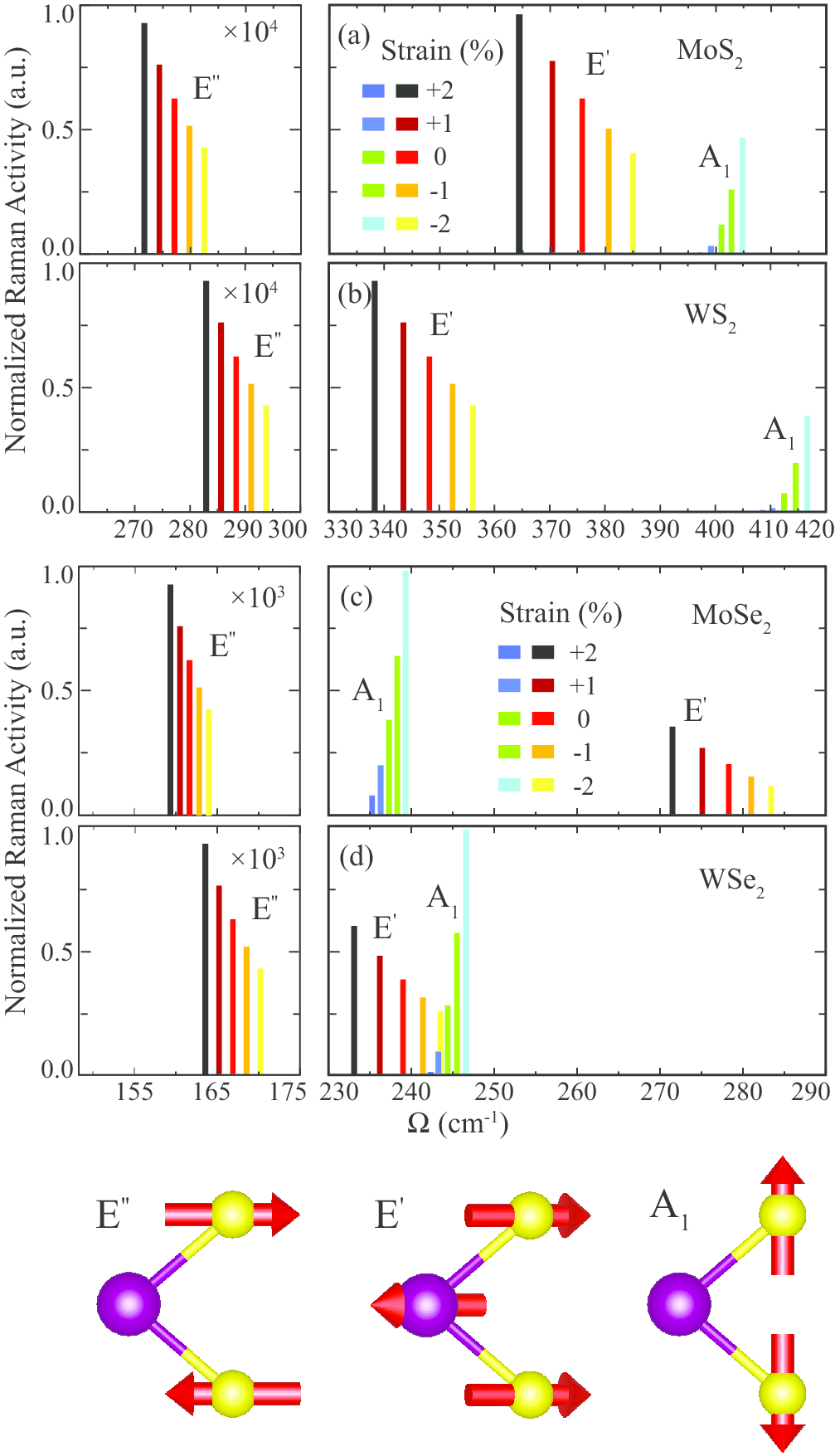}
\caption{\label{structure5}
(color line) The response of Raman active modes to the applied 
biaxial strain for single-layer (a) MoS$_2$, (b) WS$_2$, (c) 
MoSe$_2$, and (d) WSe$_2$. The vibrational motions of atoms in corresponding 
phonon modes are shown below.}
\end{figure}

\begin{figure}
\includegraphics[width=15cm]{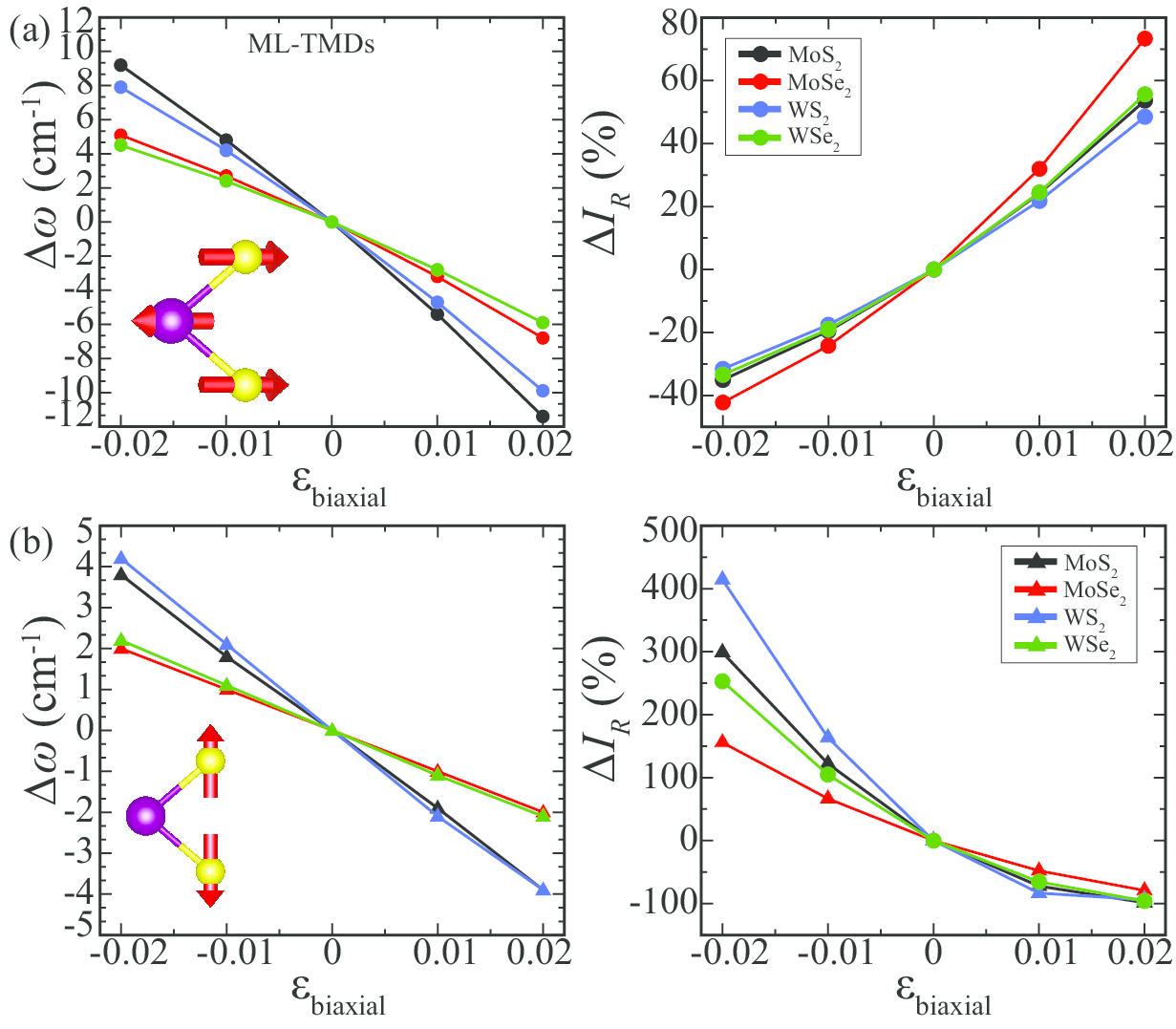}
\caption{\label{structure6}
(color line) The response of peak positions of Raman active modes and their 
Raman 
activities to the applied 
biaxial strain for single-layers of TMDs for (a) E$^{'}$ mode and (b) A$_1$.}
\end{figure}

\subsection{TMDs Single-Layer Crystals}\label{tmd}

The single-layers of TMDs has a hexagonally packed crystal structure in 
which 
the sub-layer of metal (M) atom is sandwiched between two sub-layers of 
chalcogen 
(X) atoms (see Fig. \ref{structure1}(d)). Most of the single-layer TMDs 
crystallize in either 1H or 1T phases. Here, we consider single-layers of Mo- 
and 
W-dichalcogenides which have 
1H crystal structure. Single-layer crystals of the TMDs exhibit 
$P\bar{6}/m\bar{2}$ 
space 
group symmetry. MX$_2$ (M=Mo or W, X=S or Se) crystals exhibit 9 phonon 
branches 6 of which are optical phonon 
branches (see Fig. \ref{phonons}(c)). The group theory analysis indicate that 
due to the $D_{3h}$ 
symmetry group there are 5 Raman active modes. The in-plane optical phonon 
modes are known as the E$^{'}$ and E$^{''}$ which are both doubly degenerate 
and Raman active. Additionally, an out-of-plane optical mode, A$_1$, is Raman 
active (see Fig. 
\ref{structure5}). Another 
out-of-plane optical mode, A$_2^{''}$, is the only Raman inactive mode for the 
single-layer TMDs.

The E$^{''}$ phonon mode is the lowest frequency 
optical mode which reveals only the vibration of the chalcogen atoms in 
opposite directions. The frequencies of E$^{''}$ are 
277.8, 162.5, 289.5, and 167.9 cm$^{-1}$ for MoS$_2$, MoSe$_2$, 
WS$_2$, and WSe$_2$, respectively. Since the mass of Se atom is much larger 
than that of S atom, phonon modes soften in MSe$_2$ 
crystals. The Raman activity of E$^{''}$ mode is much smaller than those of 
E$^{'}$ and A$_1$ modes in each 
single-layer. In MS$_2$ crystals its Raman activity is in the order of 
10$^{-4}$ while in MSe$_2$ structures the activity 
is in the order of 10$^{-3}$. Thus, it is not a prominent Raman peak in 
the experiments.

The effect of uniaxial strain on the vibrational spectrum of single-layer 
MoS$_2$ was investigated by Doratoj \textit{et al.} and due to the in-plane 
symmetry breaking in the crystal, the phonon-splitting was showed for the 
in-plane E$^{'}$ mode which is a direct observation of induced strain in the 
structure\cite{doratoj}. In the present study, we try to show to detection of 
strain under the biaxial case. When the crystals are biaxially stretched, 
E$^{''}$ displays 
phonon softening and demonstrates a 
phonon hardening under compressive strains as shown in Fig. 
\ref{structure5}. The corresponding mode Gruneissen parameters are given in 
Table \ref{main3}. 
Since the Raman activity of E$^{''}$ is very small, it is 
still not prominent even at high tensile strains. Because of the insignificant 
changes in the Raman 
activity of E$^{''}$, other two 
prominent Raman peaks are discussed which may determine the strain in 
single-layer TMDs.

The E$^{'}$ mode demonstrates the vibration of the transition metal atom in 
opposite direction to the chalcogen atoms. The frequencies of E$^{'}$ are 
375.8, 278.3, 348.2, and 239.0 cm$^{-1}$ for MoS$_2$, MoSe$_2$, 
WS$_2$, and WSe$_2$, respectively. The Raman activity of E$^{'}$ indicates that 
it is a prominent peak clearly observed in 
experiments\cite{Dong}. The response of the frequency of E$^{'}$ peak to the 
biaxial strain is discussed through the 
calculated mode Gruneissen parameter for each single-layer. As given in Table 
\ref{main3}, $\gamma$ values are 0.68 (0.65)\cite{gru1}, 0.53, 0.64, and 
0.54 for MoS$_2$, MoSe$_2$, 
WS$_2$, and WSe$_2$, respectively. The Raman activity of E$^{'}$ peak 
demonstrates a 
similar response to the biaxial strain with that of in monoatomic and diatomic 
cases. Due to opposite responses of E$^{'}$ and A$_1$ peaks, discussion of 
their 
relative 
Raman activities is more meaningful. 

As mentioned above, the A$_1$ phonon mode is the only 
Raman 
active 
out-of-plane optical mode. In all the single-layer TMDs the frequency of A$_1$ 
mode is higher than that of E$^{'}$ 
except for MoSe$_2$. The frequencies of A$_1$ are 401.0, 237.3, 412.4, and 
244.4 cm$^{-1}$ for single-layer MoS$_2$, 
MoSe$_2$, 
WS$_2$, and WSe$_2$, respectively. The higher frequencies in MS$_2$ crystals 
are due to smaller vertical distance of 
S-S atoms. In the unstrained single-layer TMDs, the Raman activity 
of A$_1$ peak is smaller than that of E$^{'}$ except for MoSe$_2$. In vdW 
layered materials, increasing number of layers strongly 
affects the activity of A$_1$ peak due to additional interlayer interaction. As 
given in Table \ref{main3}, $\gamma$ values for 
the A$_1$ peak are smaller than the values for E$^{'}$ which means frequency of 
A$_1$ is less affected by the in-plane strain. This is meaningful since A$_1$ 
represents the out-of-plane vibration of the atoms. The values for A$_1$ are 
0.23 (0.21\cite{gru1}), 0.21, 
0.25, and 0.22, respectively. In contrast to response of 
Raman activity of E$^{'}$ to applied strain, the activity of A$_1$ increases 
when the structure is compressed which is a result of increasing dipole length 
between vibrating chalcogen atoms when structure 
is compressed in the in-plane directions.

Although, the Raman activity of both prominent peaks changes under biaxial 
strain, it is meaningful to discuss their relative 
ratios to identify the strain in the crystal. The ratios, 
$\frac{I_{E'}}{I_{A_1}}$, 
in the unstrained structures are 5.35, 0.53, 8.32, and 1.37 
for MoS$_2$, MoSe$_2$, 
WS$_2$, and WSe$_2$, respectively. Since A$_1$ and E$^{'}$ demonstrate opposite 
responses to the biaxial strain, the ratio gets much higher when the structure 
in stretched. This is an important point 
for the identification of biaxial strain on the crystal. For the 
maximum compression (-2\%) the ratios are, 0.87, 0.12, 1.11, and 
0.26, respectively. Contrary to compressive strain, the values are enhanced 
under tensile strain case which is a strong indication of the stretched 
crystal. In the case of MS$_2$ crystals, $\frac{I_{E'}}{I_{A_1}}$ increases to 
574 and 204 for MoS$_2$ and WS$_2$, respectively. High values of the activity 
ratio can be 
clearly observed in a Raman experiment in which the samples are under tensile 
biaxial strain. 

\subsection{Anisotropic Single-Layer Crystals}

\subsubsection{\textbf{Rhenium Dichalcogenides (ReS$_2$ and ReSe$_2$) 
Single-Layer Crystals}}\label{rex}

Besides the perfect hexagonal lattice, there are also 
in-plane anisotropic single-layer crystals such as ReS$_{2}$, ReSe$_{2}$ and bp.
As shown in Fig. \ref{structure1}(e), single-layers of ReS$_{2}$ and 
ReSe$_{2}$ have  
distorted 1T (1T$^{\prime}$) crystal structure, which belongs to space group of 
\textit{P}$\bar{1}$.\cite{tongay}
Unit cell of 1T$^{\prime}$ phase consists of 
8 chalcogen atoms coordinated around 
diamond-like Re$_4$ cluster which is formed by the strong interaction between 
Re atoms. 
The angle between the in-plane unit 
cell vectors is 61.1$^{\circ}$ due to 
distortion in the crystal structure. 
 
For single-layer ReX$_2$ there are 36 phonon modes as shown in Fig. 
\ref{phonons}(d). Due to the distorted and 
anisotropic 
crystal structure, all of 
the Raman active phonon modes are non-degenerate. The 18 of the 36 phonon 
modes for both crystals are known to be Raman active from the group 
theory\cite{Feng}. 

The 18 
Raman 
active phonon modes of single-layer ReS$_2$ are classified as A$_g$-like 
(representing out-of-plane motion of atoms), 
E$_g$-like (representing 
in-plane motion of atoms), and the coupled vibrations of the atoms in both 
directions. 
There are 4 A$_g$-like modes 2 of 
which 
contain the motion of Re atoms while 
the other 2 contain motion of S atoms. The modes with frequencies of 
132.6 and 139.9 cm$^{-1}$ represent the A$_g$-like modes of Re atoms while 
the modes at 429.0 and 402.7 cm$^{-1}$ represent that of S atoms. The Raman 
activities of 
A$_g$-like modes of Re are smaller than those for the modes of S 
atoms which is because of the smaller dipole length between Re atoms. The 
frequencies 
and the Raman activities of 
A$_g$-like modes of both atoms demonstrate similar behavior under applied 
strain as in other TMDs discussed in Sec. \ref{tmd}. However, the change of 
activity of A$_g$-like modes of Re is much smaller than that 
of S atoms because at the strength of applied strain the length of Re-Re dipole 
is still small due to their strong interaction.

\begin{figure}
\includegraphics[width=15cm]{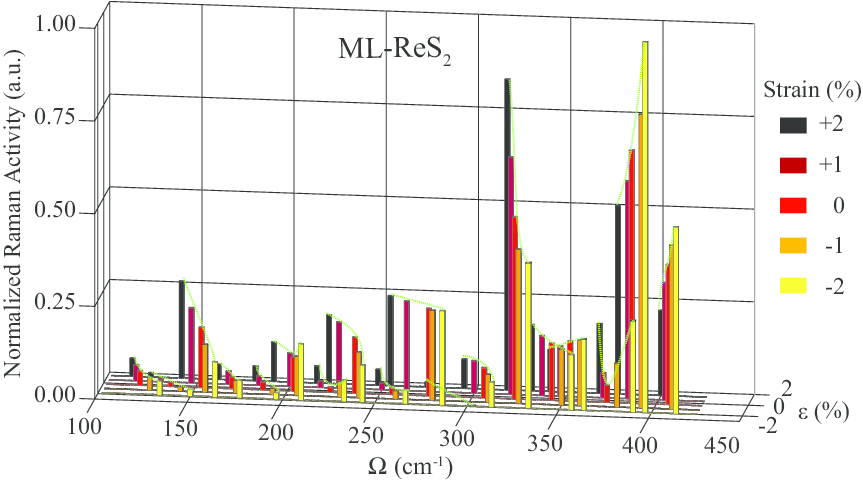}
\caption{\label{structure7}
(color line) The evolution of peak frequency of 18 Raman 
active phonon 
modes and their corresponding Raman activities under biaxial strain in 
single-layer ReS$_2$. Dashed green lines display the shift of the peak 
frequencies.}
\end{figure}

The total number of E$_g$-like modes are 6 and 4 of the modes reveal the 
in-plane motion of Re 
atoms while the other 2 demonstrate that of S atoms. The frequencies indicate 
that the motion of Re atoms occur 
at lower frequencies than those of 
S atoms. The frequencies of E$_g$-like modes of Re atoms are 151.3, 165.0, 
218.9, and 239.5 
cm$^{-1}$ and 298.2 and 307.0 
cm$^{-1}$ 
for those of S atoms. The response of the in-plane modes to the applied strain 
for both peak positions and Raman activities agree 
with those for isotropic TMDs (see the dashed green lines in Fig. 
\ref{structure7}) 

The remaining 8 Raman active modes represent the coupled vibration through 
in-plane and out-of-plane directions. 6 of the 
modes represent the vibration of only S atoms while in the other 2 modes 
coupled vibrations of Re and S atoms occur. Since for the responses of Raman 
activities of A$_g$-like and E$_g$-like modes an opposite trend is 
seen, the changes of the activity of 
coupled modes is found to be smaller. The change of Raman activity is 
determined by the 
prepotency of vibrations of atoms that is for example, if the in-plane motion 
is dominant to out-of-plane than we see an increase in the activity 
under stretching of the crystal. The most significant 
change occurs for the coupled mode 
with frequency of 412.0 cm$^{-1}$. In the mode, the out-of-plane vibrational 
motion is dominant to 
that of in-plane and thus, the activity decreases under tensile 
strain at a rate of 36\% while it increases at a rate of 47\% under compression.

In the case of single-layer ReSe$_2$, the frequencies of the Raman active 
modes significantly soften when compared with those of ReS$_2$ due to lower 
in-plane stiffness of the crystal. The highest Raman active mode has a 
frequency of 290.7 cm$^{-1}$. The A$_g$-like 
phonon modes of 
Re 
atoms have the 
frequencies of 106.5 and 116.0 cm$^{-1}$ while the frequencies are 157.0 and 
176.2 
cm$^{-1}$ for that of Se atoms. The most significant response of the Raman 
activity is found 
for the 
most 
intense peak at 260.9 cm$^{-1}$. Under compressive strain, the activity 
of the coupled mode of Re and Se atoms increases about 50\% of its unstrained 
value while it decreases 
about 55\% under tensile biaxial strain. Although, in some of the Raman active 
modes display significant changes under strain, 
due to the rigidity of the ReX$_2$ crystals they are almost irresponsive to the 
applied strain. 

\begin{figure}
\includegraphics[width=15cm]{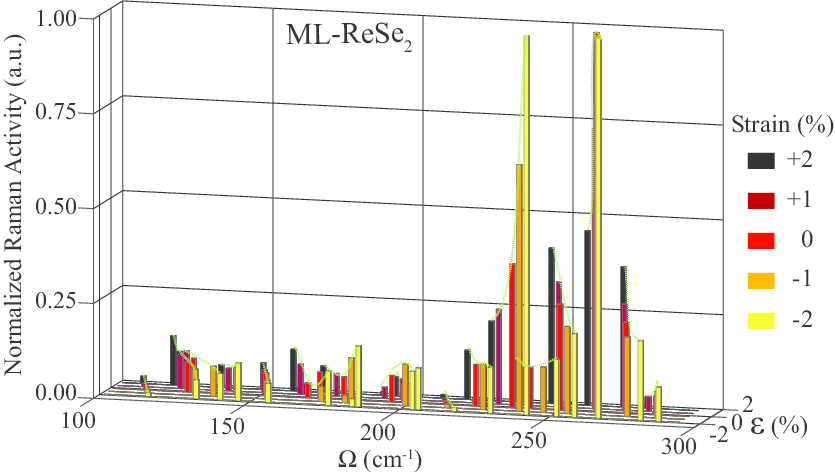}
\caption{\label{structure8}
(color line) The evolution of peak frequency of 18 Raman 
active phonon 
modes and their corresponding Raman activities under biaxial strain in 
single-layer ReSe$_2$. Dashed green lines display the shift of the peak 
frequencies.}
\end{figure}

\subsubsection{\textbf{Black Phosphorus Single-Layer Crystal}}\label{bp}

Single-layer bp is a recently synthesized, in-plane anisotropic 
member of 2D single-layer family. It is known to posses 
remarkable in-plane anisotropic electrical, optical and phonon 
properties\cite{3,4,5}. There are 4 
P atoms 
in its rectangular primitive unit cell and its crystal structure belongs to  
$Cmca$ space group. 
 
Phonon-band structure of single-layer bp demonstrates that 12 phonon branches 
exist (see Fig. \ref{phonons}(d)). 
According to the group 
theory analysis, it exhibits 6 Raman active 
phonon modes known as B$_{1g}$, B$_{3g}^1$, A$_g^1$, B$_{3g}^2$, 
B$_{2g}$, and A$_g^2$. The Raman active modes, 
B$_{1g}$, B$_{3g}^1$, and B$_{2g}$, represent the in-plane vibration 
of P atoms 
while B$_{3g}^2$, A$_g^1$, and A$_g^2$ represent out-of-plane motion. 
Experimental measurements revealed that 
only 3 of the 6 Raman active modes (A$_g^1$, B$_{2g}$, and A$_g^2$) exhibit 
prominent 
Raman intensity\cite{bp-1,bp-2}. The frequencies of the 3 prominent 
peaks are 352.1, 
424.2, and 455.2 cm$^{-1}$ for A$_g^1$, B$_{2g}$, and A$_g^2$, respectively. 
For the unstrained single-layer bp, Raman 
activities demonstrate that the A$_g^1$ and A$_g^2$ have much higher activities 
than that of B$_{2g}$ which was also observed in the 
experiment\cite{bp-1,bp-2}. 

\begin{figure}
\includegraphics[width=15cm]{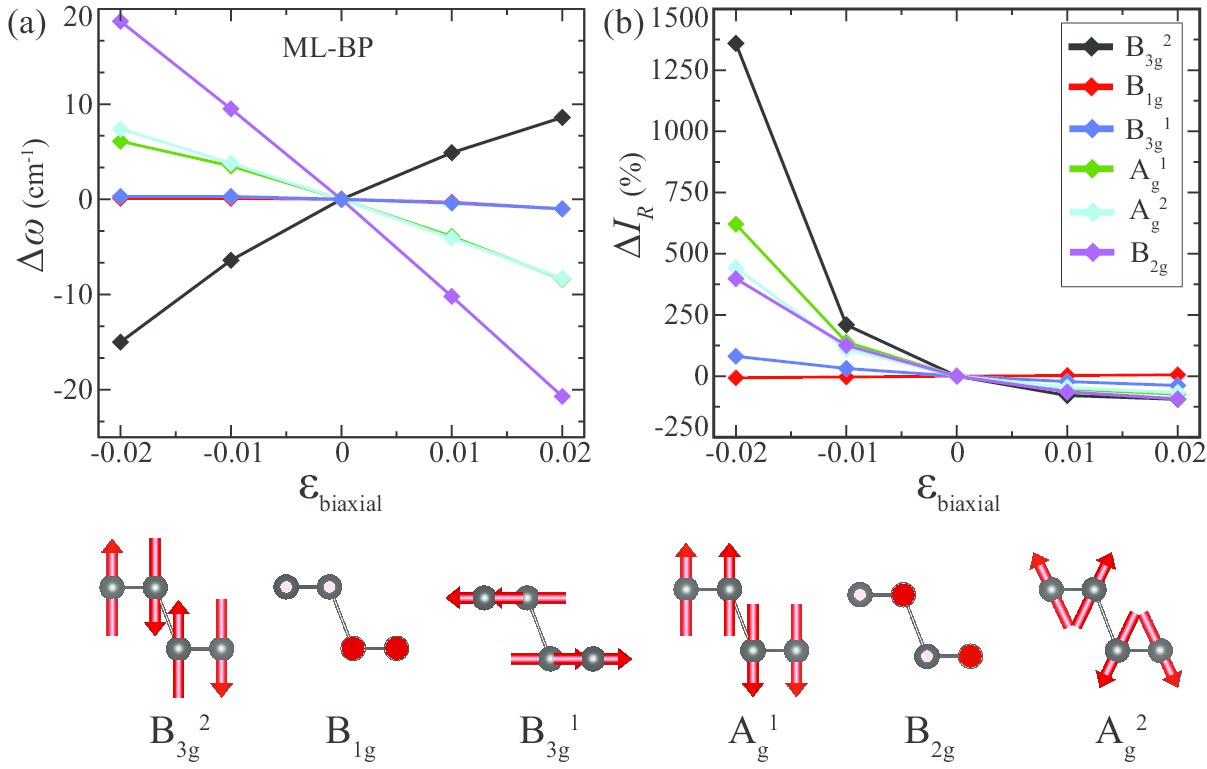}
\caption{\label{structure9}
(color line) (a) The response of frequencies and (b) Raman activities of six 
Raman 
active modes of single-layer bp to the biaxial strain. 
Vibrational motion of P atoms in each phonon mode are given below.}
\end{figure}

%%%%%%%%%%%%%%%%%%%%%%%%%%%%%%%%%%%%%%%%%%%%%%%%%%%%%%%%%%%%%%%%%%%%%%%%%%%%%%%%
\begin{table}
\caption{\label{main4} Unstrained peak frequency, $\omega_0$, the rate of 
change of peak frequency, 
$\frac{1}{\omega}\frac{d\omega}{d\varepsilon}$, and corresponding 
mode 
Gruneissen parameter, $\gamma$, for 
3 high frequency prominent peaks in single-layer bp.}
\begin{tabular}{lcccccccccccccccc}
\hline\hline
 & $\omega_0$ & $\frac{1}{\omega}\frac{d\omega}{d\varepsilon}$ & 
$\gamma$ & 
\\
 &(cm$^{-1}$) & (\%) & $-$ & 
\\\hline
A$_g^1$ &  352.1 & 1.05 & 0.53  \\

B$_{2g}$ &  559.6 & 2.32 & 1.16  \\

A$_g^2$ &  296.8 & 0.87 & 0.44  \\

\hline\hline 
\end{tabular}
\end{table}
%%%%%%%%%%%%%%%%%%%%%%%%%%%%%%%%%%%%%%%%%%%%%%%%%%%%%%%%%%%%%%%%%%%%%%%%%%%%%%%%

\begin{figure}
\includegraphics[width=15cm]{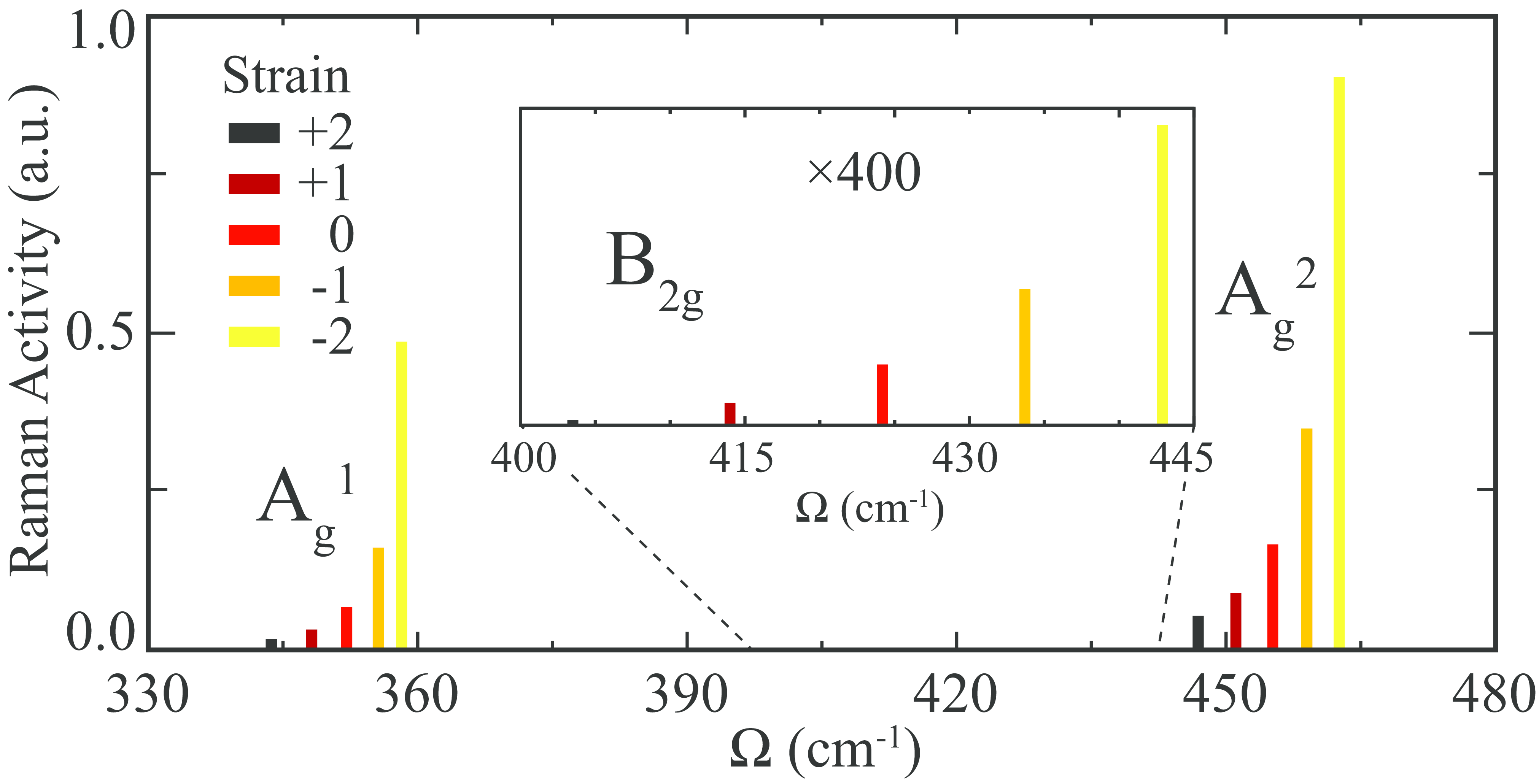}
\caption{\label{structure10}
(color line) The evolution of peak frequency of 3 prominent Raman 
active phonon 
modes and their corresponding Raman activities under biaxial strain in 
single-layer bp.}
\end{figure}

It was shown by Fei \textit{et al.} that the frequencies of vibrational modes 
of and their Raman scattering peaks in black phosphorus exhibit substantial and 
distinct shifts according to the strength of applied uiaxial 
strain\cite{fei-hakem}. Here we investigate response of the frequencies and 
Raman signals to the applied biaxial strain. When single-layer bp is biaxially 
compressed up to 2\%, the frequency of 
A$_g^2$ 
phonon mode displays a hardening from 455.2 to 462.6 cm$^{-1}$ while a 
softening 
to 446.9 cm$^{-1}$ is found 
under 2\% tensile biaxial strain. The frequency 
of B$_{2g}$ phonon mode displays the same trend as the A$_g^2$ mode under 
biaxial strain. A softening to 403.5 cm$^{-1}$ and a hardening to 442.9 
cm$^{-1}$ are seen for the frequency of B$_{2g}$ phonon mode under tensile and 
compressive strains, respectively. For the strain range of 
$\pm$2\%, the response of frequency of B$_{2g}$ is much larger than that of 
A$_g^2$ mode. Another 
characteristic Raman active mode of single-layer bp is A$_g^1$ which represents 
the out-of-plane vibrations of the P atoms in upper and lower sublayers 
in opposite directions. The frequency of the mode 
softens 
to 343.7 cm$^{-1}$ under 2\% of tensile strain while it hardens to 358.2 
cm$^{-1}$ under that of 
compressive strain. The responses of peak frequencies of the 3 prominent modes 
to 
the applied in-plane biaxial strain are compared 
through their mode Gruneissen parameters given in Table \ref{main3}. The 
greater $\gamma$ is calculated for B$_{2g}$ mode which 
exactly has the in-plane vibrational characteristic. The $\gamma$ values of 
A$_g^1$ and A$_g^2$ are smaller than that of 
B$_{2g}$ since they both demonstrate the out-of-plane vibration of P atoms.

The Raman activity of A$_g^2$ 
phonon mode is highly affected by the applied strain. The Raman activity of 
A$_g^2$ increases up to 5.4 (440\%) of its unstrained value 
under 2\% of compressive strain. However, its Raman activity is 
less sensitive to tensile strain when compared with that of compressive strain. 
Under 2\% of tensile strain, the Raman activity decreases about 68\% of its 
initial value which is much lower than that of compressive value. In addition, 
the 
Raman activity of B$_{2g}$ phonon mode is approximately 375 times smaller than 
that of A$_g^2$ mode for the unstrained 
crystal structure. The Raman activity of B$_{2g}$ displays a symmetric 
response under compressive and tensile biaxial strains. Due to the out-of-plane 
nature of the 
mode, the same trend is also illustrated for B$_{2g}$ mode. The Raman activity 
of A$_g^1$ phonon mode is in 
the order of that of A$_g^2$ mode in unstrained structure. Since the two modes 
display the same trend 
under biaxial strain, it is meaningful to compare the Raman activity of 
A$_g^1$ with its unstrained value. When 2\% of compressive strain is applied, 
the Raman activity of A$_g^1$ 
increases about 7 times of its unstrained value. 

As mentioned above, totally 6 Raman active modes exist for single-layer bp 3 of 
which have very low Raman activities (at the order of 10$^{-5}$ of the value of 
A$_g^2$ mode). The frequencies of B$_{1g}$ and B$_{3g}^1$ are 
193.8 and 232.8 cm$^{-1}$, respectively. We find that 
both the frequencies and 
the corresponding Raman activities are mostly insensitive to the applied 
biaxial 
strain. The 
reason is the vibrational characteristic of the modes. It is seen that 
the modes represent in-plane vibration of P atoms. Differing from B$_{2g}$, in 
B$_{1g}$ and B$_{3g}^1$ P atoms located at the same layer vibrate in the same 
direction. As the 
biaxial strain is applied, the thickness of 
the layer decreases but the out-of-plane symmetry of the vibrating atoms is 
mostly conserved. Thus, both peak positions and 
Raman activities mostly remain unaffected by the applied strain.

\section{Conclusions}\label{Conc}

In this study, the first-order off-resonant Raman spectra of 2D single-layers 
of 
monoatomic (graphene, Si, and Ge), diatomic (\textit{h}-BN, 
\textit{h}-AlN, and \textit{h}-GaN), in-plane isotropic TMDs (MoS$_2$, 
MoSe$_2$, WS$_2$, 
and 
WSe$_2$), and in-plane anisotropic crystals (ReS$_2$, ReSe$_2$, and bp) and 
their strain-dependent behaviors were investigated by performing DFT-based 
calculations. Our results well fit into 
the 
reported experimental results for the first-order off-resonant Raman 
activities. In addition, the effect of biaxial strain on Raman spectra of 
the single-layer crystals was analyzed in terms of their peak frequencies 
and corresponding 
Raman activities. Our findings can be summarized as follows; (i) strain can 
be directly observed 
in Raman scattering experiments by the knowledge of the peak positions of Raman 
active phonon modes, (ii) the A-peak of the single-layer Si and 
Ge 
disappear 
under sufficient tensile strain, 
(iii) especially in mono and diatomic 
single-layers, the shift of the peak frequencies is 
stronger indication of the strain rather than the change in Raman activities, 
(iv) in the case of isotropic single-layer TMDs (MoS$_2$, MoSe$_2$, WS$_2$, 
and 
WSe$_2$) the activity ratio of E$^{'}$ to A$_1$ phonon mode, 
$\frac{I_{E'}}{I_{A_1}}$, is a key for the determination of the induced strain 
since the ratio significantly increases when a tensile 
strain is applied while it decreases under compressive strain due to the 
opposite responses of the phonon modes, and (v) 
finally, 
a remarkable point for the 
anisotropic single-layers of ReX$_2$ is that there is no significant change in 
Raman 
activities under biaxial strain. 

In general, it was confirmed by the calculations that to extract
strain information in novel single-layer 2D 
crystals, peak positions of lattice vibrational modes and the corresponding 
Raman activities are 
useful.

\begin{acknowledgments}
 Computational resources were 
provided by TUBITAK ULAKBIM, High Performance and Grid Computing Center 
(TR-Grid 
e-Infrastructure). H.S. acknowledges financial support from the 
Scientific and Technological Research Council of Turkey (TUBITAK) under the 
project number 116C073.

\end{acknowledgments}

\end{document}